\documentclass[10pt,conference]{IEEEtran} 
\usepackage{cite}
\usepackage{amsmath,amssymb,amsfonts}
\usepackage{algorithmic}
\usepackage{graphicx}
\usepackage{textcomp}
\usepackage{xcolor}
\usepackage{url}
\usepackage{pdfpages}
\usepackage{subcaption}
\usepackage{svg}
\usepackage[ruled,vlined, linesnumbered]{algorithm2e}
\usepackage[inline]{enumitem}
\usepackage{scalerel}
\usepackage{tikz}
\usetikzlibrary{svg.path}

\definecolor{orcidlogocol}{HTML}{A6CE39}
\tikzset{
  orcidlogo/.pic={
    \fill[orcidlogocol] svg{M256,128c0,70.7-57.3,128-128,128C57.3,256,0,198.7,0,128C0,57.3,57.3,0,128,0C198.7,0,256,57.3,256,128z};
    \fill[white] svg{M86.3,186.2H70.9V79.1h15.4v48.4V186.2z}
                 svg{M108.9,79.1h41.6c39.6,0,57,28.3,57,53.6c0,27.5-21.5,53.6-56.8,53.6h-41.8V79.1z M124.3,172.4h24.5c34.9,0,42.9-26.5,42.9-39.7c0-21.5-13.7-39.7-43.7-39.7h-23.7V172.4z}
                 svg{M88.7,56.8c0,5.5-4.5,10.1-10.1,10.1c-5.6,0-10.1-4.6-10.1-10.1c0-5.6,4.5-10.1,10.1-10.1C84.2,46.7,88.7,51.3,88.7,56.8z};
  }
}
\newcommand\orcidicon[1]{\href{https://orcid.org/#1}{\mbox{\scalerel*{
\begin{tikzpicture}[yscale=-1,transform shape]
\pic{orcidlogo};
\end{tikzpicture}
}{|}}}}
\usepackage[hidelinks]{hyperref}
\def\BibTeX{{\rm B\kern-.05em{\sc i\kern-.025em b}\kern-.08em
    T\kern-.1667em\lower.7ex\hbox{E}\kern-.125emX}}

\IEEEoverridecommandlockouts\IEEEpubid{\makebox[\columnwidth]{ 978-1-6654-3540-6/22~\copyright~2022 IEEE \hfill} \hspace{\columnsep}\makebox[\columnwidth]{ }}

\begin{document}
\raggedbottom

\title{Scale-friendly In-network Coordination}

\author{\IEEEauthorblockN{Stefanos Sagkriotis \orcidicon{0000-0001-9438-3636}}
\IEEEauthorblockA{\textit{School of Computing Science} \\
\textit{University of Glasgow}\\
Glasgow, UK}

\and
\IEEEauthorblockN{Dimitrios Pezaros \orcidicon{0000-0003-0939-378X}}
\IEEEauthorblockA{\textit{School of Computing Science} \\
\textit{University of Glasgow}\\
Glasgow, UK}}


\maketitle

\begin{abstract}
The programmability of modern network devices has led to innovative research in the area of in-network computing, i.e., offloading certain computations to the programmable data plane. Key-value stores, which offer coordination services for many large-scale data centres, benefited from this technological advancement. Previous research reduced the response latency of key-value requests by half through deploying the store in the programmable data plane. In this work, we identify previous design decisions that have led to increased traffic generation and latency for in-network coordination services. We have developed a new in-network key-value store platform that maintains strong consistency and fault-tolerance, while improving performance and scalability over the state-of-the-art. We have designed and implemented the platform in P4, and analysed the optimisations that unlock these performance improvements. Our evaluation shows a reduction of up to orders of magnitude in latency and significant improvements in throughput. We obtain up to nine times higher throughput for scenarios with multiple participating nodes, indicative of the superior scalability the platform can offer.
\end{abstract}


\section{Introduction}

Key Value Stores (KVSs) are among the most pervasive applications in data centres. They have an instrumental role in providing configuration management, locking mechanisms \cite{Highly_available_transactions}, and web-service related operations for many large scale data centres (e.g., Google \cite{GoogleF1}, Facebook \cite{Facebook_TAO}). Previous works have focused on enhancing the performance of replicated storage by offloading computations to programmable switches, effectively utilising their line-rate processing speed to accelerate Key-Value (KV) operations. This endeavour presents a high potential in terms of performance improvements in KV response times and throughput. \par
NetChain \cite{NetChain} is one of the most prominent works in this area. The suggested KV platform can generate query responses for KV pairs placed within register arrays of programmable switches. This concept allows sub Round-Trip Time (RTT) responses to queries, enabling an order of magnitude increase in throughput and substantially lower latency than legacy KV platforms, like ZooKeeper \cite{zookeeper}. Operating entirely in-network removes need to reach a coordination server, which would require more hops, and thus the message path required for a response is reduced in half. 
\par
Other important works use Programmable Data Plane (PDP) to offload certain KV processes, like conflict detection \cite{Flair}. Their evaluation showed noteworthy performance improvements over the legacy deployment methods. More specifically, Harmonia \cite{Harmonia}, through performing in-network conflict detection for KV requests, managed to achieve higher throughput than Redis and MongoDB. The authors compared Harmonia's throughput against NetChain showing that NetChain offers orders of magnitude higher throughput than Harmonia \cite{Harmonia}. Similarly, Flair performs in-network conflict detection and redirects queries to computing nodes which would also increase hops and latency. This evidences the advantages of generating responses directly from network devices and completely omitting transactions with coordination servers. \par
NetChain remains \textbf{the fastest} in-network KVS solution to date. The tendency of switch manufacturers to increase the size of fast-access TCAM memory coming with the switches, rendered in-network KVS relevant and the best performing option. However, because of it's design it presents some limitations that make it less appealing for large-scale deployments. The employed Chain Replication (CR) mechanism in order to maintain per-item consistency among the participating nodes requires full chain traversals to fetch values from the appointed reference node \cite{Chain_replication}. This results in generating packets that require multiple hops between switches to fetch a value. \par
In this paper we propose NetCRAQ, a novel in-network KV platform which improves average read throughput by approx. $130\%$ compared to NetChain. NetCRAQ utilises the registers of programmable switches to place KV pairs and generate sub-RTT replies to queries. We leverage the line-rate packet processing features of programmable switches to implement a more complex ingress control logic that is able to decrease the average response time by reducing unnecessary pings to the tail of the chain. We demonstrate that a more complex packet processing procedure favours performance when compared to extensive header parsing. Our design preserves the necessary information for all KVS operations within the data plane, minimising interactions with the Control Plane (CP). \par
We reassess the scalability achieved by NetChain and show that increasing chain lengths are deteriorating its attainable latency and throughput. We promote higher scalability and offer lower average latency over increasing chain lengths and higher average throughput: up to $9.46\times$ higher throughput for a chain of 8 nodes, and $4$ orders of magnitude lower latency. The routing mechanism is designed to achieve high parsing efficiency and low overhead over the underlying transfer protocol. NetCRAQ operates under strong consistency, while the replication method can be adapted to work with relaxed consistency in favour of performance, something not previously available. \par
Overall, this paper contributes by:
\begin{itemize}
    \item Identifying weaknesses and performance limitations of KV platforms that operate in PDP.
    \item Discussing the impact of design decisions in PDP environments. 
    \item Proposing a new in-network KV platform with major performance and scalability improvements over the state-of-the-art. 
    \item Providing the accessory routing and processing mechanisms to efficiently deliver strong consistency and flexibility. 
\end{itemize} \par

\section{KVS Design for the Data Plane} \label{background} 
Research in the workload characteristics of deployed KVS shows that they accommodate read-mostly workloads: the read-write ratio is 380:1 for Google F1 \cite{GoogleF1}, and 500:1 for Facebook TAO \cite{Facebook_TAO}. Primary-backup variations, like CR, are specifically designed for read-mostly workloads, making them an appealing candidate for the underlying replication method of data centre KVSs. 


\subsection{Chain Replication} 
\begin{figure}
    \centering
    \begin{subfigure}[t]{.49\columnwidth}
        \includegraphics[scale=.269]{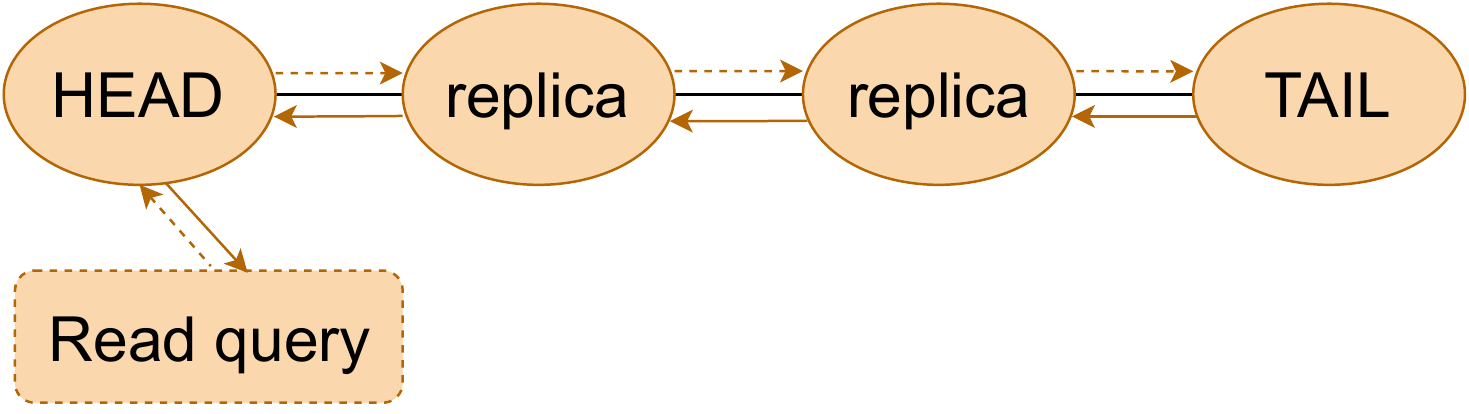}
        \caption{Message path for a read query in CR.}
        \label{CR_read_path}
    \end{subfigure}
    \begin{subfigure}[t]{.49\columnwidth}
        \includegraphics[scale=.269]{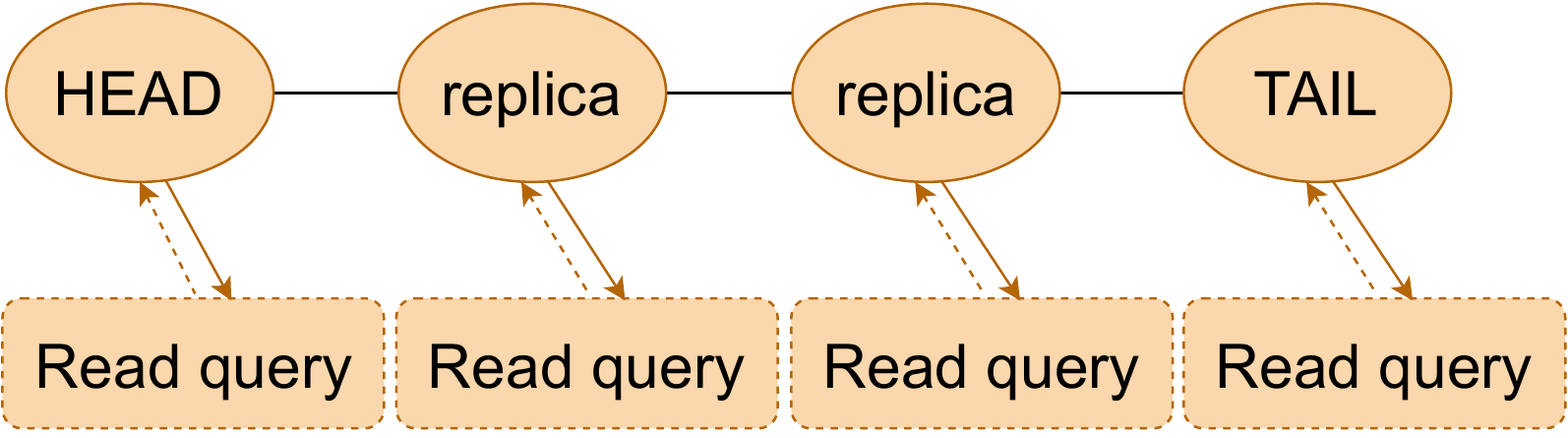}
        \caption{Message path for read queries to clean objects in CRAQ.}
        \label{CRAQ_read_path}
    \end{subfigure}
    \\
    \begin{subfigure}[t]{.49\columnwidth}
        \includegraphics[scale=0.269]{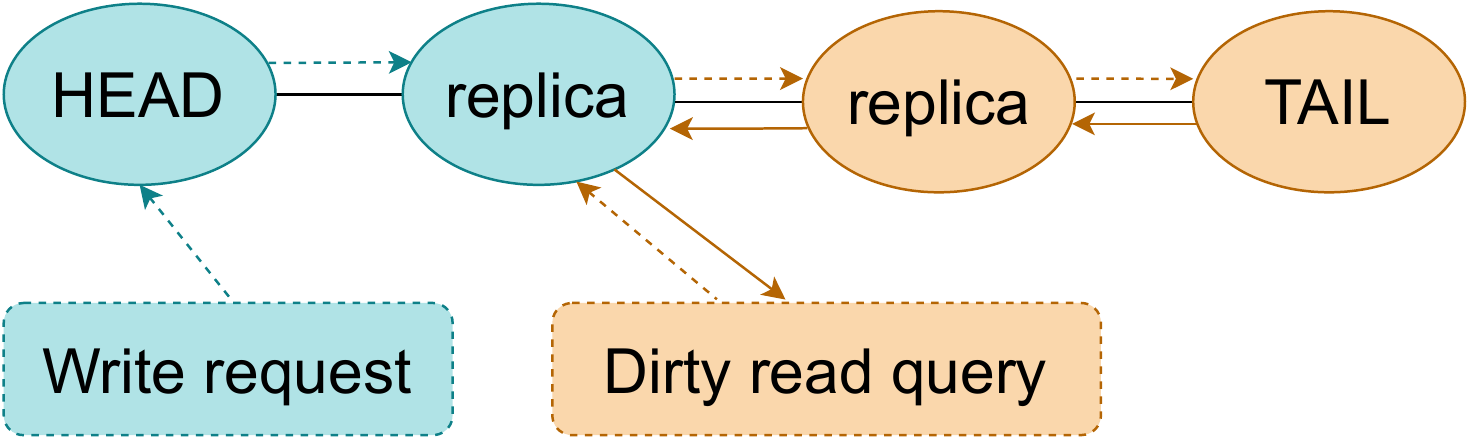}
        \caption{Message path for read queries to dirty objects in CRAQ.}
        \label{CRAQ_dirty_read_path}
    \end{subfigure}
    \caption{Comparison of message path for a read query in CR vs CRAQ.}
\end{figure}

Under CR, the participating nodes/programmable switches form a chain and each has a distinct role: head, tail, or replica. All of the participating nodes hold the same KV pairs. Write queries originate from the head and then propagate across all replica nodes until they reach the tail. The tail issues a response which acts as an acknowledgement for the write. A tail node is also responsible for responding to read queries. Only the tail is considered to be up-to-date with the latest commit for a value and acts as a reference point for the entire chain. In Figure \ref{CR_read_path}, we show the path of a read query (dashed arrows) and its response (solid arrows). Replicas can replace the head or the tail in case of a failure.  \par
Defining the tail as the reference point allows per-key consistency for the entire chain to be achieved. When a write query reaches the tail, it has certainly been processed by all previous chain nodes. Therefore, all chain nodes are updated with its latest version. If the write query is lost before reaching the tail, then all subsequent reads will be replied with the previous version for this object. This ensures consistency in replies. \par

\subsection{In-network KVS} 
\subsubsection{IncBricks}\label{IncBricks} 
The IncBricks \cite{IncBricks} paper was among the first to capture the potential of PDP devices because of their location in a query's message path. The message path for a query in a legacy KVS would start from the client, go through network devices, and finally it would reach the servers that host the KVS. IncBricks substituted the need to reach coordination servers for some types of queries and suggested a shorter path for cached values which stays between the client and the network devices. 
Network programmability was achieved with the use of programmable network processors which offered limited instructions and were tied to proprietary programming tools provided by the manufacturer. Despite this limitation, the realisation of this new role for network devices greatly reduced the query response time and increased the attainable throughput over legacy KVS deployments. \par
\subsubsection{NetChain}\label{NetChain} 
This new role of switches was further explored by NetChain, which by using P4 \cite{P4_16} managed to deploy an in-network KVS in high-performance ASICs (instead of Network Processor Units (NPUs)), and therefore achieved greater performance than IncBricks. \par
The query response mechanism that was employed relies heavily on the incoming packets that are processed using the match-action pipeline \cite{PSA}. A custom packet format was used, layered over UDP, which contained the following fields: {\fontfamily{cmtt}\selectfont OP} - the type of operation (read, write), {\fontfamily{cmtt}\selectfont KEY} - the ID of the object in question, {\fontfamily{cmtt}\selectfont VALUE} - its value, {\fontfamily{cmtt}\selectfont SEQ} - a monotonically increasing sequence number that mitigates out-of-order deliveries, {\fontfamily{cmtt}\selectfont SC} - the number of chain nodes in the header, {\fontfamily{cmtt}\selectfont S\textsubscript{k}} - IP of the k\textsuperscript{th} participating node. Storing the IPs of the nodes in the header aims at reducing the amount of stored data per switch and allowing dynamic mapping of data to chains. \par
We notice that the suggested packet structure can add a significant amount of overhead bytes, especially for bigger chains where all the participating node IPs have to be added in the header. For a 4-node chain, NetChain's header is 58 bytes, and grows by 32 bits with every node addition. The linear growth of the packet size with the chain length can cause increased parsing times and add complexity every time fields are added or removed \cite{P8}, which according to the platform's design happens each time a query is processed. In a data centre topology the overhead computations result in scalability loses and wasted resources. This design choice forces the administrator to choose between performance and redundancy. \par
Another issue arises from the use of a monotonically increasing value in the {\fontfamily{cmtt}\selectfont SEQ} packet field, which is 16 bits by default. This size allows just $65,536$ operations before the field overflows, effectively limiting the amount of writes to this number.  \par
NetChain's replication method for the data plane is based on a variation of CR. The reasoning behind this choice reflects the limitations and features of the deployment environment and provides important lessons. Firstly, CR has small redundancy requirements to achieve fault tolerance: to survive $f$ node failures, it requires $f+1$ nodes. This is an important parameter when considering that it translates to the amount of programmable switches in use. Secondly, CR presents low implementation complexity by requiring a simple commit \& forward pipeline to execute a write query among the chain nodes. \par
We observe some performance-limiting factors. Based on the principle that only the tail can reply to read queries, the amount of generated packets is substantial: for $n$ participating nodes, $2n$ packets are required for read queries and $n + 1$ for write queries. Considering the workloads that the platform needs to sustain by using the aforementioned design, we notice the following limitations: \begin{enumerate*}
  \item generating messages for the tail results in high packet gain for the platform. It requires network resources for extensive parsing and forwarding;
  \item the chain's reply rate is limited to the throughput that can be provided by the tail node, being the only one responsible to reply. This heavily harms scalability;
  \item directing all queries to a certain node can also be a root cause for hot-spots within the topology;
  \item the response latency increases linearly with the chain length because of the increasing number of hops.
\end{enumerate*}\par

\subsection{CRAQ} 
Chain Replication with Apportioned Queries (CRAQ) \cite{CRAQ}, employs a different design but operates in a similar manner to CR: the nodes formulate a chain and each node can be a head, tail, or replica. The key differences with CR are: CRAQ's ability to distribute load across all chain nodes -- effectively enhancing scalability, and its ability to operate under relaxed consistency guarantees to benefit performance. \par
In CRAQ, each KV pair can be either clean, in which case there are no pending commits for its value, or dirty, which means that there is a most recent commit which is yet to be acknowledged by the tail. Therefore, multiple versions of a value might correspond to a key. CRAQ places this information inside each participating node. Upon receiving a read query, each node can either: respond to it, if the version is clean (cf. Figure \ref{CRAQ_read_path}), or redirect the query to the tail in order to fetch the latest version (cf. Figure \ref{CRAQ_dirty_read_path}). Writes operate similarly to CR: a node has to propagate a write down the chain until it reaches the tail and then be acknowledged as the latest clean version. Once this happens, the rest of the chain nodes are notified and can delete previous versions of this object's value. \par

\section{NetCRAQ Design} \label{overview}
\begin{figure}
    \centerline{\includegraphics[width=0.35\textwidth]{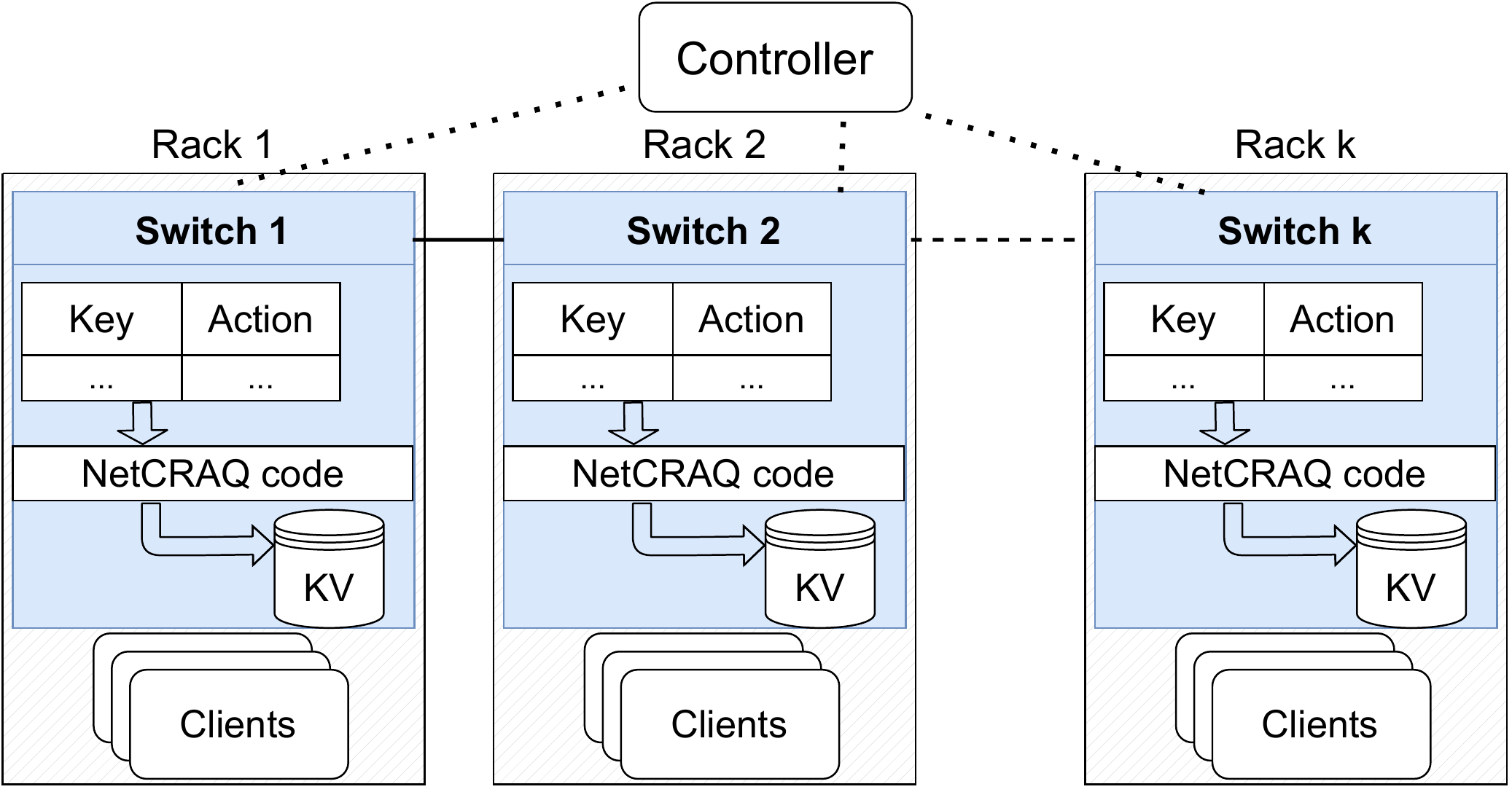}}
    \caption{Overview of NetCRAQ.}
    \label{NetCRAQ_overview}
\end{figure}

NetCRAQ's design delivers a fault-tolerant, in-network KVS with focus on high throughput and scalability, minimum packet gain, and strong consistency. The design allocates functionality between control and data plane according to the advantages and disadvantages of each. Time-critical computations are placed in the data plane to utilise line-rate performance. The CP is responsible for network-wide operations, like failure detection and recovery. An overview of the design is shown in Figure \ref{NetCRAQ_overview}. NetCRAQ supports multi-level topologies and is able to direct queries to nodes using IP forwarding. This could be utilised to formulate different chains within a topology, using the available resources in the most efficient way. \par

\subsection{NetCRAQ Data Plane Design} \label{dp_design} 
NetCRAQ's data plane design is responsible for two main operations: storing/retrieving the KV pairs in the relevant data structures and processing and forwarding queries and coordination messages. We present our design for these operations. \par

\subsubsection{Implicit KV state in PDP}
A key difference between CR and CRAQ is the way that new writes are processed. In CRAQ, for each object $k$, there are potentially multiple versions, $n$. In CR, appending multiple values for an object is not required and instead the only ``clean'' version exists in the tail. In the context of programmable switches, to satisfy CRAQ's requirement, $n$ register cells need to be available to commit writes. For this reason, the switch is initialised with $k\times n$ sequential register cells reserved, forming an array. We call this the {\fontfamily{cmtt}\selectfont objects\_store} array. 
Each object consumes $n$ number of cells in the array to allow dirty commits to be appended at the end of the last commit. 

The state of each object (clean/dirty) has to be retrieved to determine the future control logic operations. We implicitly define the state as clean iff the latest committed value exists in the first cell of the object's space within the array. This implicit definition is based on the principle that every previous value is deleted upon a successful write of an object, i.e., when this is acknowledged by the tail node. This design choice enabled fast packet processing without the need to divert packets to the CP or use extensive parsing. 


\subsubsection{Packet format}
Since extensive packet parsing cannot be avoided when messages have to traverse the entire chain, even by employing CRAQ, truncating the packet size and reducing packet modifications on each switch should enable faster forwarding between the participating nodes. NetCRAQ's packet format follows a simpler approach by having just three fields layered over UDP:
\begin{itemize}
    \setlength\itemsep{0.5mm}
    \item {\fontfamily{cmtt}\selectfont KV\_OP}: defines  the  type  of  the  operation:  read  request/reply,  write  request, acknowledgement.  (2 bit)
    \item {\fontfamily{cmtt}\selectfont KEY\_ID}: contains the key id.  (32 bit)
    \item {\fontfamily{cmtt}\selectfont VALUE}: the field containing the value for the specific key.  (128 bit)
\end{itemize} \par
The number of participating nodes and the IPs thereof are omitted from the packet and instead placed within the switch, thus reducing the parsing time for all KV operations and coordination messages. The CP is responsible for updating the roles according to changes, instead of relying on the incoming packets. 

\subsubsection{Control logic} 
\begin{algorithm}
\footnotesize
\SetAlgoLined
    objects\_store = register[$k*n$]\;
     read\_index = register[$n$]\;
     write\_index = register[$n$]\;
\uIf{kv\_op == READ}{
    get\_read\_index(KEY\_ID)\;
    get\_my\_role()\;
    \uIf{meta.read\_index == 0}{
        clean\_read(KEY\_ID)\;
        generate\_reply();\
    }
    \uElseIf{meta.my\_role == TAIL}{
        dirty\_read(KEY\_ID)\;
        generate\_reply();\
    }
    \uElse{
        forward\_to\_tail()\;
    }
}
\uElseIf{kv\_op == WRITE}{
    get\_write\_index(KEY\_ID)\;
    get\_my\_role()\;
    \uIf{meta.write\_index == 0}{
        clean\_write(KEY\_ID)\;
        forward\_to\_tail()\;
    }
    \uElse{
        \uIf{meta.write\_index $>$= NUMBER\_OF\_VERSIONS}{
            drop()\;
        }
        \uElse{
            dirty\_write(KEY\_ID)\;
             forward\_to\_tail()\;
        }
    }
    \uIf{meta.my\_role == TAIL}{
        clean\_write(KEY\_ID)\;
        generate\_acknowledgement()\;
        multicast()\;
    }
}
\uElseIf{kv\_op == ACKNOWLEDGEMENT}{
    clean\_write(KEY\_ID)\;
}
\caption{Control Logic} \label{control_logic}
\end{algorithm}
The control logic entails all the necessary operations for interacting with the KV pairs and managing the network traffic. All KVS operations are atomic to protect the values from simultaneous accesses. Values obtained from parsing the NetCRAQ header are matched against a pre-defined table that dictates the action that is executed when a match occurs. \par 
Metadata fields, used for branching decisions, are filled by the CP in advance. These fields contain values that need to be regularly retrieved to manage incoming traffic. For example, the role of each switch or the IP of the switch appointed to be the tail of the chain. This is a design difference with NetChain that enables faster overall parsing and forwarding by retrieving information from fast-access memory instead of packet parsing. \par
The control of packets that contain the NetCRAQ header is primarily dictated by the {\fontfamily{cmtt}\selectfont KV\_OP} field. The allowed operations for this field are: {\fontfamily{cmtt}\selectfont READ}, {\fontfamily{cmtt}\selectfont READ\_REPLY}, {\fontfamily{cmtt}\selectfont WRITE}, and {\fontfamily{cmtt}\selectfont ACKNOWLEDGE}. Deletes happen in the form of a {\fontfamily{cmtt}\selectfont WRITE} operation, since the memory is statically managed and cannot be freed upon removal of KV pairs. Each operation is designated by a different value in this field. \par
Algorithm \ref{control_logic} shows the complete control logic. If the identified operation is a {\fontfamily{cmtt}\selectfont READ}, the next decision is based on the position of the value within the register. If a value exists in the first position of the object's register space, we know that the version is clean, otherwise it is dirty. The next stage includes checking the role of the node. Only a tail node can reply to a read with a dirty version, while the rest can only reply with clean versions of an object. Writes in the tail node may be committed but not yet acknowledged by all nodes, therefore replying with the latest value is not voiding consistency. \par
For {\fontfamily{cmtt}\selectfont WRITE} operations, we use a similar technique to identify clean writes: the index for the next available cell to commit a write should be at the first cell of the object's register space. Otherwise, the write should be considered dirty and, once committed, should be forwarded to the tail. If the write attempts to exceed the object's predefined register space, it is considered to be out of bounds and the packet is dropped. All write requests that arrive at the tail are considered to be the latest clean ones and acknowledgements are generated for the rest of the chain. To quickly and efficiently update the rest of the chain, we use the {\fontfamily{cmtt}\selectfont multicast} functionality of P4, which automatically generates the correct amount of necessary packets, based on the size of the chain and transmits them at once. \par
Lastly, receiving an {\fontfamily{cmtt}\selectfont ACKNOWLEDGEMENT} message in any node implies that the contained value of the message is the latest clean version for the object. Therefore, all previous versions are deleted and the relevant indices for the object are reset.

\subsection{NetCRAQ Control Plane Design} \label{cp_design}
The CP is responsible for installing all the match-action rules related to forwarding, KV operations, and failure recovery. 
The CP allocates a different IP in each switch. This IP is stored within the metadata of each switch and determines whether a query will be replied on the arriving node or forwarded. We use the IP protocol for this and modify the header accordingly to forward to tail or generate acknowledgements. This provides flexibility for multi-level topologies, offering integration with load-balancing protocols, like Equal-Cost Multi-Path. Furthermore, the CP determines and allocates the roles of the switches within the chain. Based on the number of participating nodes and the distance between them, different role allocation techniques can be used for a more flexible deployment. For example, the CP can integrate the {\fontfamily{cmtt}\selectfont meter} extern, offered by P4, to identify potential hot spots within the topology and re-adjust the chain lengths and the KV pairs within each register.



\subsection{Handling Failures} \label{fail_recovery}
Failure mitigation happens in two phases: 
\begin{enumerate*}
    \item immediate redirection of traffic to a failover node to reduce the traffic loss; and
    \item complete recovery with a replacement node and re-installation of forwarding rules and KV pairs.
\end{enumerate*} \par
When a node remains unresponsive for a certain amount of time, the client can automatically direct requests to a different chain node. This time can be adjusted based on what is considered as a prolonged lack of response according to the average response rate of the network. Once the failure is noticed by the CP, the forwarding rules are updated by removing the node from the forwarding tables and the multicast group. \par
In the second phase, a new node re-enters the chain. To maintain consistency, we follow CRAQ's approach to identify which node will be used to copy KV pairs from. The CP, depending on the position of the failed node, decides the node that will be used to copy the KV pairs to the new node. Due to space limitations, we refer the reader to the original CRAQ paper for the complete list of scenarios \cite{CRAQ}. The recovery node remains offline while the CP copies the KV pairs from an online node. During this phase, the CP also disables any writes across the chain in order to preserve consistency. When the copy is complete, the node is added in the forwarding tables and the multicast group of the chain. 

\section{Evaluation} \label{evaluation}
NetChain compared against legacy KVS and proved the clear performance benefits from generating responses in sub-RTT distance using fast data plane memory accesses \cite{NetChain}. NetChain's findings allow us to \textbf{safely infer} that, if NetCRAQ's performance is equal or better than NetChain, then it is also faster than legacy methods. Therefore, we directly compare NetCRAQ's performance against NetChain in a series of tests concerned with: throughput, latency, mixed workloads, and scalability. \par
Our testbed runs a bare-metal installation of Ubuntu 18.04 (kernel: 4.15.0-140-generic) on Intel Core i7-4790 CPU and 16GB of DDR3 RAM. P4 behaviour is emulated using the reference BMv2 switch \cite{BMv2} - compiled using performance flags. The topology is generated and managed using Mininet \cite{Mininet} and P4-utils \cite{p4-utils}. The CP is written in Python and communicates with the Mininet switches using Thrift \cite{thriftAPI} and the P4-utils API.

\subsection{Throughput}
We evaluate the throughput of both platforms based on the maximum attainable rate at which they can provide responses to queries. The measurements are in Queries Per Second (QPS). We direct millions of packets to each switch while increasing the packet rate. The maximum attainable response rate is considered the rate at which the response rate starts to decrease and the response latency rises. \par
\begin{figure}
    \centerline{\includegraphics[scale=0.37]{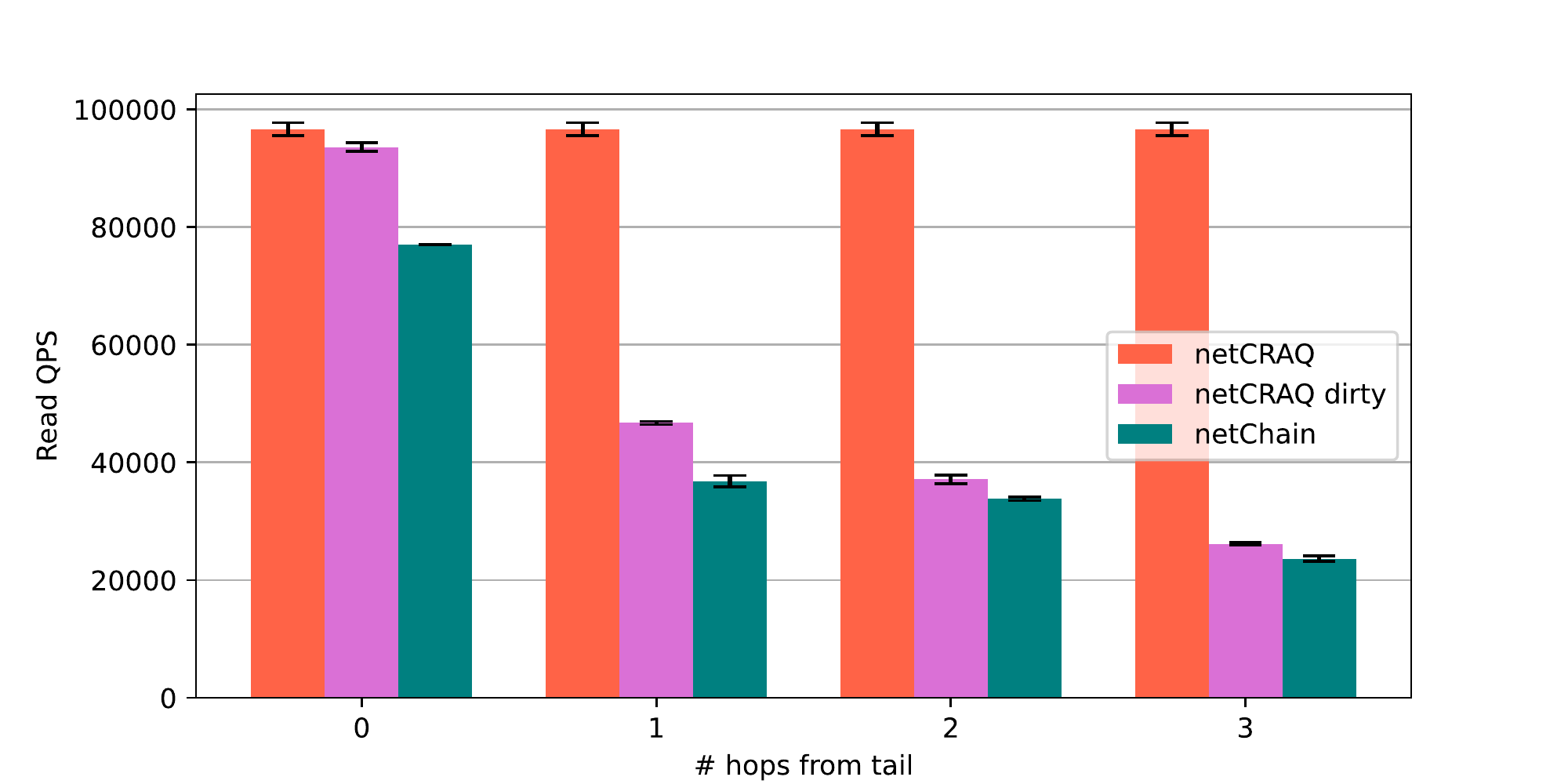}}
    \caption{Max read QPS vs distance from tail.}
    \label{QPS_vs_distance}
\end{figure}

In figure \ref{QPS_vs_distance}, we test the impact of NetCRAQ's ability to provide responses to read queries of a clean version. We monitor the throughput each node is able to achieve given the distance it has from the tail. NetCRAQ's throughput appears unaffected by distance when the queried object is clean. The reduction in required hops and computations creates a big performance difference in favour of NetCRAQ: $4.08\times$ higher throughput for queries directed to the head of the chain. In case of dirty objects, throughput is still higher than NetChain with the difference being attributed to the smaller packet size used by NetCRAQ, 72 overhead bytes for NetChain vs 20 bytes for NetCRAQ. This difference results in smaller parsing times, which when the number of hops increases is less apparent. When dirty queries are generated directly at the tail, the amount of processing required to generate a reply is the only factor impacting performance since hops are minimum. In that case NetCRAQ shows 22\% higher throughput than NetChain, proving higher overall computation efficiency. When queries are directed to the head, the amount of nodes between the source of the queries and the tail is introducing limitations in throughput. Here NetCRAQ is 10.5\% faster. \par

\begin{figure}
    \centerline{\includegraphics[scale=0.37]{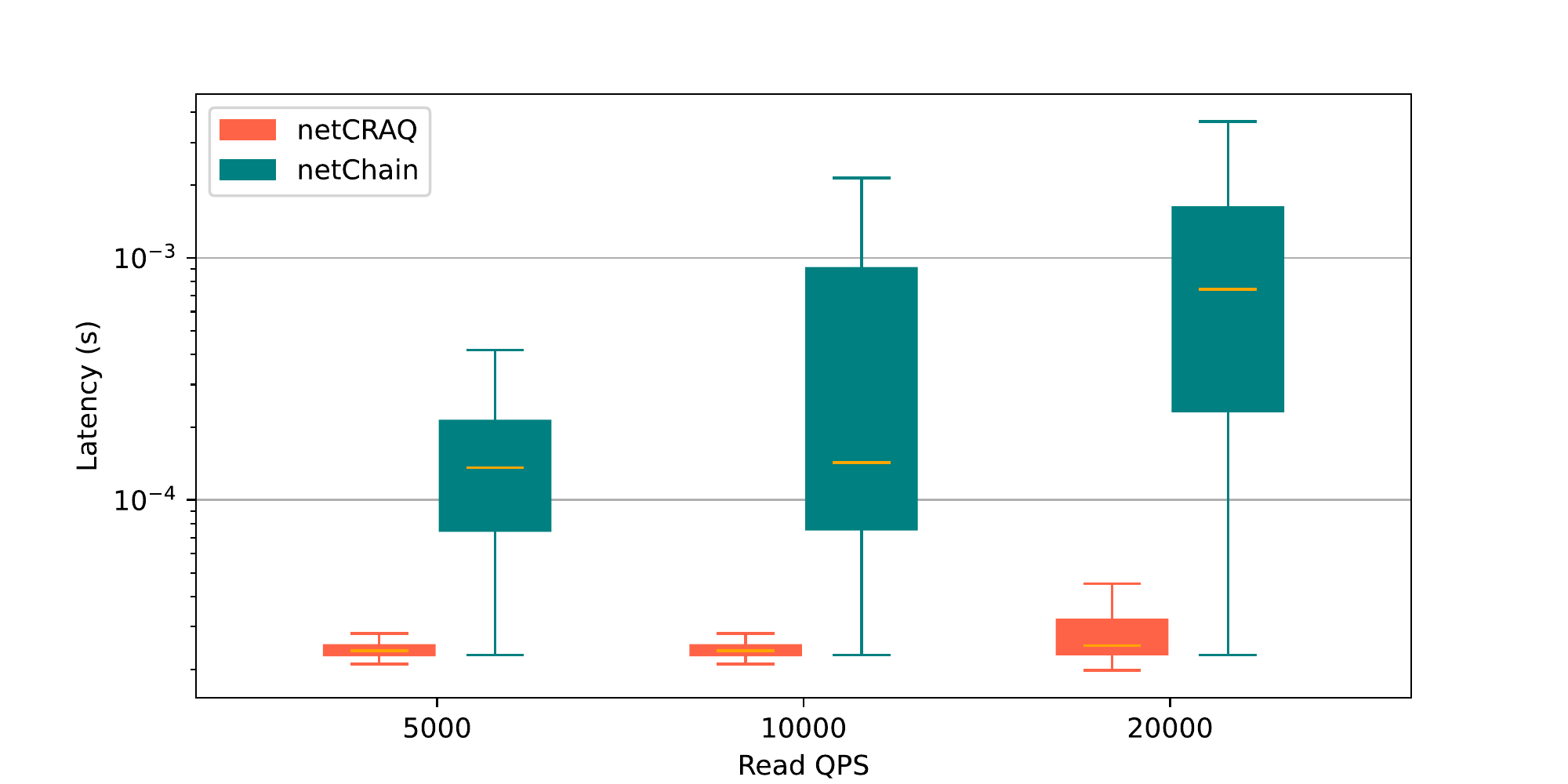}}
    \caption{Response latency vs QPS.}
    \label{Read_latency}
\end{figure}


\subsection{Latency}
As the rate of queries rises, sustaining the same response latency becomes increasingly harder with continuous atomic operations causing contention of the available resources. To investigate the latency of the two platforms under different loads, we use the same 4-node chain and generate an increasing number of read queries. Figure \ref{Read_latency} displays the obtained measurements for this scenario. In the displayed measurements, we include metrics from all participating nodes, regardless of distance from tail. Providing consistent latency with different chain lengths adds flexibility to the platform and the ability to adapt to the requirements of the KVS and the network. NetCRAQ shows a steady latency response that rises marginally with the number of read queries. NetChain presents a big variance in response latency which is related to the varying distance from tail. The difference becomes more significant as the number of QPS rises: two orders of magnitude faster responses for 5k and 10k QPS, and three orders of magnitude for 20k QPS.


\subsection{Mixed workloads}
\begin{figure}
    \centerline{\includegraphics[scale=0.37]{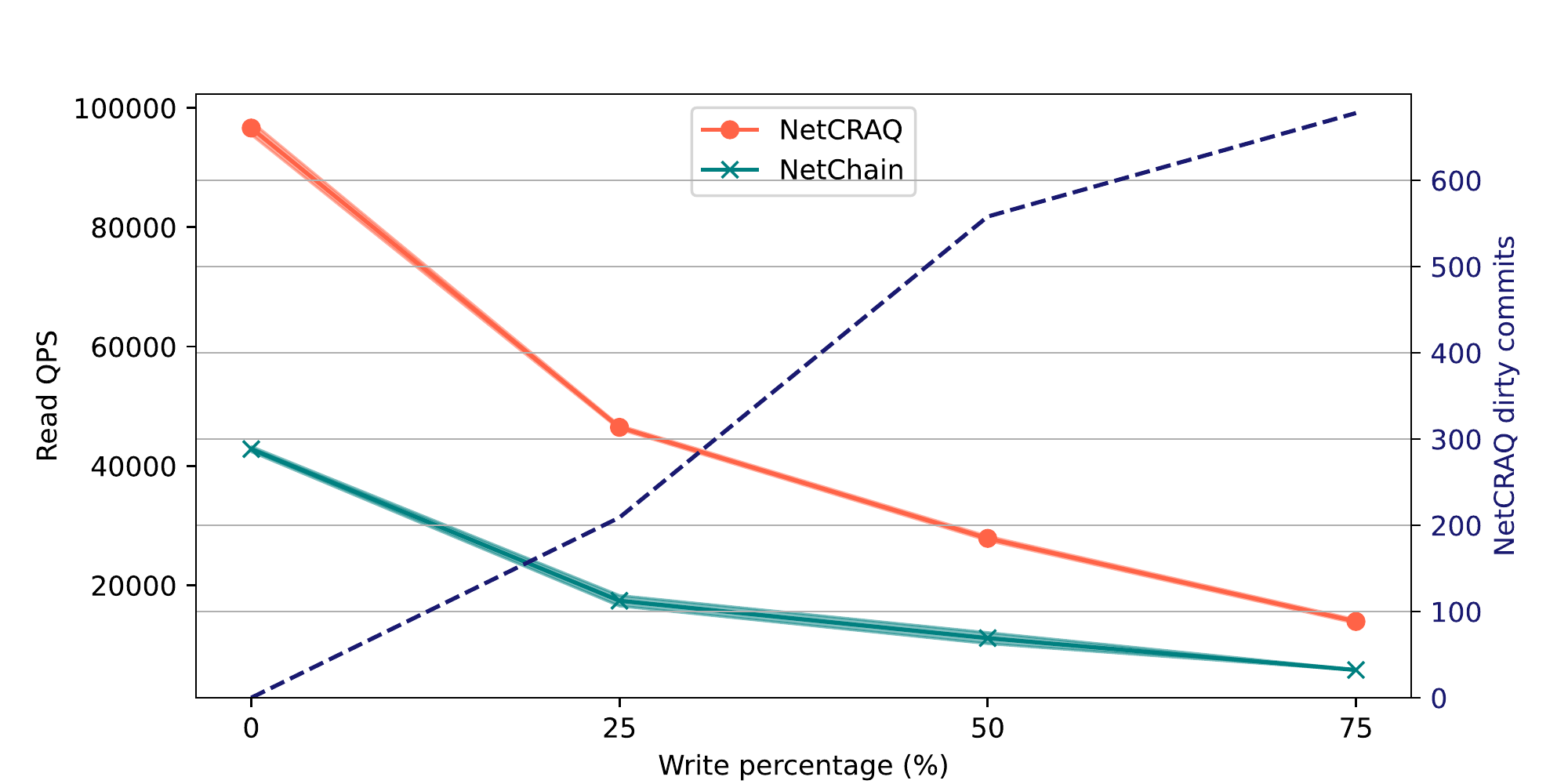}}
    \caption{Performance under mixed read/write workloads.}
    \label{mixed}
\end{figure}

We evaluate the platforms under realistic workloads containing a mix of reads and writes. The results are shown in Figure \ref{mixed}. Starting from a read-only workload, we gradually increase the percentage of writes with a step of 25\%. The performance of the platforms is judged by their attainable response rate. NetCRAQ achieves more than double the read throughput for all write percentages. Its read efficiency enables higher throughput regardless of the write percentage.
Adequate register cells need to be budgeted to maintain all dirty versions before they can be acknowledged by the tail. This is depicted by the increasing amount of dirty commits as write percentage rises, observed in the right y axis of Figure \ref{mixed}.
\begin{figure}
    \centerline{\includegraphics[scale=0.37]{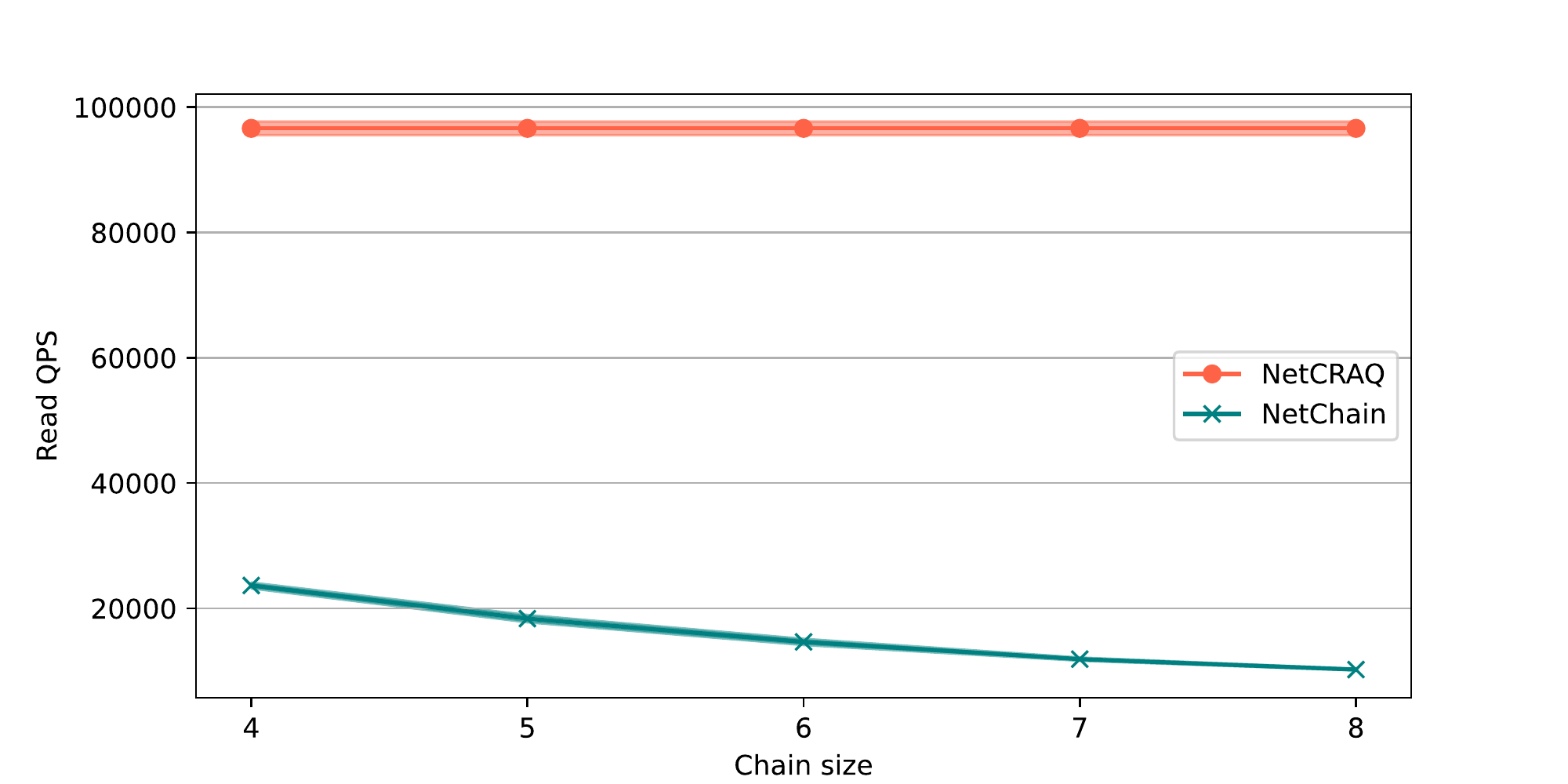}}
    \caption{Read throughput vs chain length.}
    \label{varying_chain}
\end{figure}
\subsection{Scalability}
We validate the scalability of both platforms in Figure \ref{varying_chain}. Here, the comparison is between read queries directed to the head of the chain. We vary the chain length from 4 to 8 nodes. 
The results showed that in a chain of 8 nodes, the throughput of NetChain was reduced in half. On that account, we did not proceed to further increase the chain length as the performance gap was already representative of the throughput losses that occur. 
The throughput difference can be up to $9.46\times$ in favour of NetCRAQ, for the case of 8 chain nodes. This difference stems mainly from the ability of nodes to respond directly to read queries, thus reducing unnecessary forwarding. By requiring only a single RTT to respond to a read query, the performance remains the same regardless of chain size. For the case of NetChain, the amount of computations necessary in order to respond grows with the number of participating nodes, hence the performance difference.

\section{Conclusion} \label{conclusion}
We study and evaluate the performance limitations of previous state-of-the-art in-network coordination platforms. We discuss the design choices that generate these performance limitations. Our design utilises fast PDP processing to offload a complex packet processing logic, minimising packet modifications and message exchanges. We suggest that high-level reconfiguration of chains and failure handling should be conducted in the CP. We present a new, faster, in-network coordination platform -- NetCRAQ. Without harming strong consistency, we effectively increase mean throughput, reduce mean latency, and improve scalability. Our design allows for reduced packet gain and alleviation of traffic hot-spots within the network.

\bibliographystyle{ieeetr}
\bibliography{Bibliography}
\end{document}